# Manifestation of sub-Rouse modes in flow at the surface of low molecular weight polystyrene

K. L. Ngai[1,2], S. Capaccioli[1,2], Daniele Prevosto[2], Luigi Grassia[3]

[1]*Dipartimento di Fisica, Università di Pisa, Largo B. Pontecorvo 3, I-56127, Pisa, Italy*

[2]*CNR-IPCF, Largo B. Pontecorvo 3, I-56127, Pisa, Italy*

[3]*The Second University of Naples, Department of Aerospace and Mechanical Engineering, Via Roma 19, 81031 Aversa (CE), Italy*

Correspondence to: K. L. Ngai, ngai@df.unipi.it

**Abstract**

The presence of a viscoelastic mechanism distinctly different from the segmental α-relaxation and the Rouse modes within the glass-rubber transition zone of polymers had been justified by theoretical considerations, and subsequently experimentally verified in different bulk polymers by various techniques, and in several laboratories. It is referred to in the literature as the sub-Rouse modes, naturally because their time-scales are longer than the segmental α-relaxation but shorter than the Rouse modes. The sub-Rouse modes were also found in polymer thin films by the creep compliance measurements of McKenna and co-workers [J. Polym. Sci., Part B: Polym. Phys. **46**, 1952 (2008).]. Apparently the mobility of the sub-Rouse modes is enhanced in thin films as evidenced by shifting to shorter times on decreasing the film thickness $h$. However, the shift of the sub-Rouse modes is less than the segmental α-relaxation, which is caused by the lesser enhancement of mobility of the former than the latter, a property explained by the Coupling Model. On reducing the film thickness $h$ of high molecular weight polystyrene, there is increasing separation of the sub-Rouse modes in time scales from the segmental α-relaxation, resulting in the decrease of the rubbery plateau observed in the creep compliance experiment. Thus, the important fact established by experiment and theoretical considerations is the enhanced mobility of sub-Rouse modes in thin PS films by the combination of effect from the free surface, finite size, and induced chain orientations, concomitant with the segmental α-relaxation. Induced chain orientations effect is present only when $h$ is less than the end-to-end distance of the high molecular weight polymer chains. In this paper, the proven enhanced mobility of sub-Rouse modes at the surface of polymers is used to explain recent experimental investigations of viscous flow at the surface of low molecular weight PS by Chai et al. [Science, **343**, 994 (2014)], and by Yang et al. [Science, **328,** 1676 (2010).]. Viscous flow of polymers is by global chain motion, therefore the observed large reduction of viscosity at the surface of low molecular weight PS originates from the sub-Rouse modes, and not the segmental α-relaxation. This distinction is not commonly recognized in the current literature. The accerleration of the sub-Rouse modes at the surface explains the experimental findings.

## 1. Introduction

Theoretical consideration as well as experimental evidences in the glass-rubber transition zone of bulk amorphous polymers have shown that the segmental α-relaxation is not followed immediately by the Rouse modes[1-7]. In between these two better known visocoelastic mechanisms are the new modes, referred to as sub-Rouse modes[3-8], with length within one chain longer than the segmental α-relaxation but shorter than the Gaussian submolecule, the basic unit needed for the formation of the Rouse modes. Review of the history leading to the discovery of these intermediate viscoelastic mechanisms in various polymers by experiments was given in Ref.[5]. The best way to separate out the contributions from the three mechanisms within the glass-rubber transition zone is by shear compliance (creep)[3,4], precision dielectric, and internal friction measurements[10-16]. The measured shear compliance $J(t)$ is rigorously the sum of the contributions from segmental α-relaxation, $\hat{J}_\alpha(t)$, the sub-Rouse modes, $\hat{J}_{sR}(t)$, and the Rouse modes, $\hat{J}_R(t)$. Experiments on various polymers including polystyrene and analyses of data have determined the extent of contributions of these three viscoelastic mechanisms[1,17,18]. A recent review has been given in Ref.[17], and here we go straight to the essential results. For entangled high molecular weight polystyrene (PS), it has been shown that $\hat{J}_\alpha(t)$ lies within the range bounded by the glassy compliance $J_g = 0.93 \times 10^{-9} \mathrm{Pa}^{-1}$ and $J_{e\alpha} \approx 4 \times 10^{-9} \mathrm{Pa}^{-1}$, i.e.,

$$J_g \leq \hat{J}_\alpha(t) \leq J_{e\alpha}. \tag{1}$$

The sub-Rouse modes contribution, $\hat{J}_{sR}(t)$, exist within the bounds

$$J_{e\alpha} \leq \hat{J}_{sR}(t) \leq J_{sR} \approx 10^{-7} \mathrm{Pa}^{-1}. \tag{2}$$

The Rouse modes contribution, $\hat{J}_R(t)$, fall within the range,

$$J_{sR} \leq \hat{J}_R(t) \leq J_{plateau}, \tag{3}$$

where $J_{\mathrm{plateau}} \approx 10^{-5}\ \mathrm{Pa}^{-1}$ is the entanglement plateau compliance of PS.

The values of $\hat{J}_{sR}(t)$ contributed by the sub-Rouse modes span over a considerable range. Hence, only the sub-Rouse modes with $\hat{J}_{sR}(t)$ closer to $J_{e\alpha} \approx 4 \times 10^{-9} \mathrm{Pa}^{-1}$ have properties closer to that of the segmental α-relaxation. The classical studies by Plazek and co-workers found thermorheological complexity of the compliance spectra and viscoelastic anomalies[5,6,18-24], which were confirmed over the years by other workers[25-31]. The cause is traced to the presence of the three viscoelastic mechanisms and the different temperature dependencies of their effective relaxation (or retardation) times, $\tau_\alpha$, $\tau_{sR}$ and $\tau_R$.[3-6,8] The segmental α-relaxation time $\tau_\alpha$ has the strongest, $\tau_{sR}$ the intermediate and $\tau_R$ the weakest temperature dependence. The segmental α-relaxation is well known to be dynamically heterogeneous involving cooperative or correlated motion of repeat units within a length-scale. The properties of the more recently discovered sub-Rouse also indicates cooperative dynamics, albeit to a lesser extent than the segmental α-relaxation[3-6,13]. Despite the fundamental nature of the findings of thermorheological complexity in the glass-rubber transition zone, no explanation has been given by in the literature except the singular one by the Coupling Model (CM)[4,8,17,24,25,32]. The explanation is based on the difference in the degrees of cooperativity of the three mechanisms. In the order of decreasing degree of cooperativity are the segmental α-relaxation, the sub-Rouse modes, and the Rouse modes. Degree of cooperativity is characterized in the CM by the coupling parameters. Hence we have the relations between the coupling parameters of the thre mechanisms, $n_\alpha, n_{sR}$, and $n_R$, given by by $n_\alpha > n_{sR} > n_R = 0$. The sub-Rouse modes with $\hat{J}_{sR}(t)$ and $\tau_{sR}$ closer to $J_{e\alpha} \approx 4 \times 10^{-9}$ Pa$^{-1}$ and $\tau_\alpha$ respectively have larger $n_{sR}$. The above is a short summary of the characteristics of the three mechanisms in the glass-rubber transition zone of *bulk* high molecular weight entangled PS.

The focus of the present work is on the viscous flow at the surface of low molecular weight PS, for which experimental measurements have been made recently[33,34]. At the free surface of supported and freestanding PS thin films, the mobility of the segmental α-relaxation is much higher than in the bulk. This was anticipated in the very first paper of applying the CM to PS thin films in 1998[35] by making the statement: "In addition, polymer chains on or near the surface will have an increased mobility due to fewer interactions with neighboring chains; i.e. half of the neighboring chains are missing at the surface. Again, this reduction of intermolecular constraints leads to a decrease of the coupling parameter. As $h$ decreases the surface to volume ratio increases and the reduction of $T_g$ becomes larger.". The emphasis on the free surface effect was repeated in 2002[36], where one can find the statement: "we pointed out that the reduction of intermolecular coupling in the film depends on the distance from the nearest surface and hence the same is true for the decrease in *n* or the resultant enhancement of local segmental mobility. The largest decrease of *n* from its bulk value occurs at the free surface and the change diminishes continuously when going towards the center of the film. This idea is consistent with the computer simulation results that the mobility near the surface is higher [22–28] and also the simplified three-layers model proposed later by Mattsson *et al.* [8].".

In high molecular weight (MW) PS film with thickness comparable or less than the end-to-end distance of the chains, there is induced chain orientation which can also reduce intermolecular coupling, but this effect does not exist in the viscous flow at at the free surface of low MW PS, the focus of the present paper. The free surface effect and the finite size effect (i.e., when thickness $h$ less than the cooperative length-scale) act together to mitigate intermolecular constraints, which corresponds in the framework of the CM to a reduction of the coupling parameter from the bulk value $n_{\alpha,bulk}$ to a smaller value $n_\alpha(h,j)$. Here $j$ is the $j$-th layer counting from the surface layer, which is the first. It was stated explicitly in the 2002 paper: "The largest reduction of $n_\alpha$ and $\tau_\alpha$ occurs at the free surface layer and monotonically become less for layers located further into the interior", and repeated verbatim in the recent 2013 paper[17].

The key equation of the CM is the dependence of the segmental α-relaxation time $\tau_\alpha$ on $n_\alpha$ given by

$$\tau_\alpha(T) = [(t_c)^{-n_\alpha}\tau_0(T)]^{1/(1-n_\alpha)}. \tag{4}$$

where $t_c$=1 to 2 ps for PS and $\tau_0$ is the primitive relaxation time with value independent of $h$ and $j$. Based on Eq.(4), the 2006 paper[37] gives a layer-by-layer description of the attenuation of the free surface effect on reduction of $n_\alpha(h,j)$ and $\tau_\alpha(h,j,T)$ when going towards the interior of the film. It can be easily verified from Eq.(1) that a smaller $n_\alpha(h,j)$ than $n_{\alpha,bulk}$ leads to a shorter $\tau_\alpha(h,j,T)$ than $\tau_{\alpha,bulk}(T)$, and hence the corresponding reduction of the bulk glass transition temperature to $T_g(h)$ found in supported and freestanding PS thin films.

Experiments in bulk polymers have shown that the sub-Rouse modes are also cooperative but to a lower degree than the segmental α-relaxation[4,13,17,25,38-40], and have smaller bulk coupling parameter, $n_{sR}$, than $n_\alpha$ of the segmental α-relaxation. At the surface, intermolecular coupling and cooperativity of the sub-Rouse modes are reduced for the same reason given for the segmental α-relaxation. Correspondingly, the coupling parameter of the sub-Rouse modes at the surface is much reduced from the bulk value $n_{sR,bulk}$. Like the segmental α-relaxation, the sub-Rouse modes are also intermolecularly cooperative, albeit to the lesser degree and with a smaller coupling parameter $n_{sR,bulk}$ than $n_{\alpha,bulk}$. The CM equation (4) is general and applicable to all cooperative processes including the sub-Rouse modes, which takes the form

$$\tau_{sR}(T) = [(t_c)^{-n_{sR}}\tau_{0,sR}(T)]^{1/(1-n_{sR})}. \tag{5}$$

For the same reason, the effects of the free surface and the finite size of thin film cause a reduction of the bulk coupling parameter $n_{sR,bulk}$ to smaller values of $n_{sR}(h,j)$. From this result and by applying Eq.(5) to both the bulk and the thin film, it can be easily verified that $\tau_{SR}(h,j)$ is shorter than $\tau_{SR,bulk}$. Naturally the smallest value of $n_{SR}(h,j)$ and the shortest $\tau_{SR}(h,j)$ are at the $j$=1 free surface layer.

In the following section we first briefly review the creep compliance experiments on nanobubble inflation freestanding PS thin films of McKenna and co-workers[41-43] and their observation of the simultaneous accerleration of the segmental α-relaxation and the sub-Rouse modes. The retardation times $\tau_\alpha$ and $\tau_{sR}$ both becomes shorter on decreasing the film thickness $h$, which will be used to address the results from recent study of surface viscosity of a low molecular weight PS by Chai et al.[33] and Yang et al.[34] Explanation of the experimental data by the sub-Rouse, that is consistent with all our previous works, is the objective of this paper. The viscous flow at the surface of low MW PS experiments provides another case of the manifestation of the sub-Rouse modes, which can be explained by the CM. In the final section before conclusion we mention other surface diffusion experiments in non-polymeric materials where huge enhancement of diffusivity was observed and explained quantitatively by the CM.

**2. Manifestation of sub-Rouse modes in the dynamics of polymer thin films**

Isothermal biaxial creep compliance, $D(t)$, of unsupported nanobubble inflated ultra-thin films of high MW polymers was measured over the glass-rubber transition zone by McKenna and coworkers[41-43]. For film of any thickness $h$, the creep curve shifts to shorter times on decreasing temperature. However, accompanying the shift is the decrease of the rubbery plateau compliance. The decrease becomes more dramatic in thinner films and at lower temperatures. An example of this anomalous behavior of creep compliance taken at 69, 72, and 75 C can be seen in Fig.1 for a 36 nm thick PS film with MW=994,000 Da and PDI=1.07.

The ranges of the additive contributions to shear compliance from the segmental α-relaxation, $\hat{J}_\alpha(t)$, the sub-Rouse modes, $\hat{J}_{sR}(t)$, and the Rouse modes, $\hat{J}_R(t)$, have been given in Eqs.(1)-(3) respectively. From the relation, $J(t)=6D(t)$, and Eq.(2). the sub-Rouse modes contributions, $\hat{D}_{sR}(t)$, to the biaxial compliance lie within the range,

$$(6.7 \times 10^{-10} \text{Pa}^{-1} \approx D_{e\alpha}) \leq \hat{D}_{sR}(t) \leq (D_{sR} \approx 1.67 \times 10^{-8} \text{Pa}^{-1}). \qquad (6)$$

The tip and the end of the arrow indicate $D_{sR}$ and $D_{e\alpha}$ respectively. The sole purpose of Fig.1 is to demonstrate the simultaneous observation of the sub-Rouse modes and the segmental α-relaxation in thin PS films. By considering the change of the biaxial compliance data on decreasing film thickness $h$, we have explained and concluded in Ref.[17] that both the sub-Rouse modes and the segmental α-relaxation are accerlerated on decreasing $h$, but to a lesser extent for the former than the latter. Here we can use Fig.1 to elucidate simply this fact. It is clear from Fig.1 that the creep compliance $D(t)$ data obtained at 72 and 75 C are contributed entirely by the sub-Rouse modes. In order for the sub-Rouse modes of the *bulk* PS with MW=994,000 Da to be seen in the experimental time window of Fig.1, the temperature has to be much higher than $T_g$=98.8 C. Since the sub-Rouse modes appear within the experimental time window at 72 and 75 C, therefore clearly the sub-Rouse modes have been acclerated by the effect of the free surface and possibly also the finite size effect in the 36 nm thick high-MW PS film.

In Fig.2 we compare the master curve $D(t)$ of the film with the master curve $J(t)$ of bulk PS of high MW=600,000 Da.[22] The slope of the log-log plots of the data in the sub-Rouse regime of the thin film is about a factor of 2 smaller than that in the bulk. This significant change of slope in thin film can be taken as evidence of faster sub-Rouse modes contributing to higher compliance are accerlerated less than slower sub-Rouse modes contributing to lower compliance. These experimental findings had been explained by the Coupling Model (CM) equations (4) and (5)[17] from the segmental α-relaxation coupling parameter $n_\alpha$ being larger than all the coupling parameters $n_{sR}$ of the sub-Rouse modes, and also the faster sub-Rouse mode contributing to larger value of $\widehat{D}_{sR}(t)$ has smaller coupling parameter $n_{sR}$. On decreasing the film thickness $h$, the coupling parameters $n_\alpha(h)$ and $n_{sR}(h)$ of all modes are reduced, but the acceleration of dynamics is much larger for all the modes with larger coupling parameter[17]. Thus the faster sub-Rouse modes lag behind the slower sub-Rouse modes, and all of them lag behind the segmental α-relaxation in their shifts to shorter times. This effect of bifurcation of the sub-Rouse modes from the segmental α-relaxation is absent in thick films like bulk, but becomes increasingly important on decreasing $h$. When the master curves of creep compliance data constructed for different film thickness $h$ are presented and compared over the same time window (see Fig.9 in Ref.41), the effect shown is the decrease of the plateau rubbery compliance on decreasing $h$. Thus the simultaneous accelerations of the sub-Rouse modes and the segmental α-relaxation but to a less degree for the former than the latter give an explanation of the decrease of the plateau rubbery compliance on thinning the film observed by McKenna and coworkers. Other details of the explanation were given before in Ref.[17]. Exactly the same as described in the above for PS was found in polycarbonate by Mckenna and co-workers[43]. Their creep compliance data of the 22 nm film in Fig.2a[41] and the master curve in Fig.2b[41] have essentially the same properties as Figs.1 and 2 herein, from which we can reach the same conclusions.

      There is an analogue of the effect found by McKenna et al. on reducing the film thickness of high MW PS. Instead of thinning the PS film, reduction of $n_\alpha$ and $n_{sR}$ in bulk PS can be achieved by dissolving PS in the solvent tri-*m*-tolyl phosphate. The presence of the

solvent increases the average separation the repeat units to mitigate the intermolecular interaction, and hence reduce the coupling parameters of all modes. Creep compliance measurements of 17% polystyrene solution[7] have shown that the segmental α-relaxation shifts to shorter times much more than the sub-Rouse modes, resulting in a much broader glass-rubber transition zone in the solution than in bulk PS. The retardation spectra of the bulk PS and its 17% solution in Fig.1 of Ref.[7] clearly demonstrate the acceleration by dissolution of the segmental α-retardation times by reduction of $n_\alpha$ is larger than the sub-Rouse modes by reduction of $n_{sR}$. Moreover from the change of the shape of the retardation spectrum, sub-Rouse mode with longer retardation time is accererated more than those with shorter retardation times, in support of the same found by McKenna and coworkers on reducing the film thickness of the PS thin films and the CM explanation[17].

## 3. Direct evidence of acceleration of sub-Rouse modes at surface of polymers from surface viscosity measurements

A novel investigation of enhanced surface mobility was reported by Chai et al.[33], using the geometry of a stepped PS film on a substrate. They measured the viscosity above and below the bulk $T_{gB}$ of the low molecular weight PS with $M_w$=3000 g/mol. Above the bulk $T_{gB}$=343 K or 70 C, the entire film is involved in viscous flow. However, below $T_{gB}$, flow occurs only in the near-surface region, made possible by the high mobility at the surface. At temperatures sufficiently far below $T_{gB}$, the flow measured comes totally at near the surface.

Before we proceed futher in considering the surface viscosity data of Chai et al., it is important to recognize that the viscoelastic creep compliance measurements[5,22,23,38,40] of PS, Selenium, and poly(methylphenylsiloxane) (PMPS), stress relaxation of poly(methyl-para-tolyl-siloxane) (PMpTS)[44], and light scattering of PMPS[38,39,45] and PMpTS[44], all of low molecular weights, show the presence of the segmental α-relaxation and the sub-Rouse modes, but not the Rouse modes, because the chains are too short to support compliance contributed by the Rouse modes[1,5]. This fact was established before[5,46] from the shear compliance data of Plazek and O'Rourke[22] for PS with low MW=3400 g/mol and $T_g \approx 70$ C. From about 100 C down to $T_g \approx 70$ C, they found the recoverable compliance, $J_r(t) = J(t) - t/\eta$, are all less than $J_{SR} \approx 10^{-7} \text{Pa}^{-1}$, and hence the data are contributed by the sub-Rouse modes and the segmental α-relaxation.[5,17,22,46] This can be seen from Fig.7 of Ref.[22], where the final increase of $J_r(t)$ is due to the presence of a higher molecular weight tail in the polydisperse sample. The same was found in monodisperse poly(methylphenylsiloxane) with low molecular weight of 5000 g/mol[38]. Thus, also in the of 3000 g/mol low molecular weight PS studied by Chai et al., at temperatures above and below the bulk $T_g$= 70 C, the only viscoelastic mechanisms present are the segmental α-relaxation and the sub-Rouse modes. Nevertheless, viscous flow of the low molecular weight PS in the bulk or at the surface is performed exclusively by the sub-Rouse modes, and not by the segmental α-relaxation. Hence, the sub-Rouse modes are exclusively the relevant viscoelastic mechanism in the experiment carried out by Chai et al. both in the bulk and at the surface. The enhancement of fluidity found is a direct proof that the sub-Rouse modes are accerlerated at the surface. This is consistent with the shift of the sub-Rouse modes to shorter times on decreasing film thickness observed by creep compliance measurements of high MW nanobubble inflated thin PS films[41-43] by McKenna et al., and expectation from the CM considerations.

Chai et al. was able to infer from the data the Arrhenius behavior of the surface viscosity at temperatures below $T_{gB}$, with activation energy $E_a \sim 337 \pm 20$ kJ mol$^{-1}$. There is another measurement of the viscosity of low MW polystyrene films on silicon at different temperatures by Yang et al.[34] which precedes Chai et al. and gives similar results. The PS used has $M_w$=2400 g/mol and PDI=1.06. They measured the viscosity of the PS films on silicon at different temperatures and found that the transition temperature for the viscosity decreases with decreasing film thickness. By analyzing the data, they deduced the presence of a highly mobile surface liquid layer, which dominates the flow in the thinnest films studied, and has Arrhenius $T$-dependence with activation energy of 185 kJ/mol. The magnitude of the change is consistent with Chai et al. Like the measurement of surface viscosity of 3000 g/mol PS by Chai et al. below $T_{gB}$, the surface viscosity deduced from their experiment on 2400 g/mol PS by Yang et al. is transpired by the sub-Rouse modes.

Previous efforts to account for the various viscoelastic measurements of bulk and thin films of polymers using the CM[4,8,17,38] have shown that sub-Rouse modes are intermolecularly coupled or cooperative, in analogy to the segmental α-relaxation. In the present case of interest on low MW PS, the sub-Rouse modes are responsible for viscous flow in the bulk and at the surface, while the segmental α-relaxation accounts for the enthalpic glass transition. At the surface, intermolecular couplings are mitigated, the degree of cooperativity of both the sub-Rouse modes and the segmental α-relaxation are reduced, and one can expect enhanced mobility of both viscoelastic mechanisms. Direct evidence of the accerleration of sub-Rouse modes, caused by the presence of the free surfaces, are found in the creep compliance data of nanobubble inflated PS thin films from the studies of McKenna et al., and examples are shown here in Figs.1 and 2. Within the context of the CM, reduction of intermolecular coupling of the sub-Rouse modes at the surface has the consequence of the sub-Rouse modes coupling parameter $n_{sR,surf}$ becomes smaller than the value $n_{sR,bulk}$ in the bulk. Eq.(5) of the CM can be used to calculate the sub-Rouse modes relaxation times at the surface and in the bulk. The much shorter $\tau_{sR,surf}$ at the surface than $\tau_{sR,bulk}$ in the bulk leads to the explanation on theoretical grounds of the experimental findings of enhanced surface flow by Chai et al. and Yang et al. An early experiment showing the surface has extraordinary high mobility was determined by Tanaka et al.[47] from the temperature dependence of lateral force at a given scanning rate as early as the year 2000. In this paper Tanaka et al. had successfully explained their data by the Coupling Model. Other works showing the importance of surface effcct include the surface nanohole recovery experiment[48,49].

Large enhanced surface diffusion has been observed in indomethacin, a small molecular glass-former[50]. It is also found in the surface of shear bands of mechanically deformed metallic glasses[51]. In both cases, the Coupling Model (CM) is able to explain[52,53] quantitatively the large enhancement of mobility at the surface from experiments.

There is no doubt that the free surface is an important cause of enhanced mobility of both the segmental α-relaxation and the sub-Rouse modes of polymers. The latter is amply demonstrated by the surface flow experimental data considered in this work. Notwithstanding, finite size effect is another contributing factor in polymer thin films when the thickness is comparable to the cooperative length-scale of the segmental α-relaxation. It acts alone in causing significant reduction of $T_g$ in systems without free surface as shown by experiments. Notable examples include the confinement of PMPS in nanocoposites[54], in nanometer glass pores of PDMS and PMPS by Schonhals and coworkers[55-77], and the study by Simon and coworkers[58].

4. Conclusion

Supported by creep compliance measurement of low molecular weight PS, we show that the sub-Rouse modes are intermoleculary coupled and cooperative, and are responsible for viscous flow in the bulk and at the free surface. The experimental observed large reduction of viscosity at the free surface is direct evidence of enhancement of mobility of the sub-Rouse modes by the mitigation of intermolecular coupling at the surface, occurring simultaneously with the same effect on the segmental α-relaxation. Previous creep compliance measurements of nano-bubble inflated PS thin films of high molecular weight have already shown evidence of enhancement of the mobility of the sub-Rouse modes. Altogether, these recent advances in the study of dynamics of polymer thin films have shown not only the change of the glass transition temperature effected by the segmental α-relaxation is interesting, but also that of the sub-Rouse modes and the entire glass-rubber transition zone. The Coupling Model had been successful in accounting for the viscoelastic anomalies caused by the breakdown of thermorheogival complexity of bulk polymers. It continues to explain the changes of the different viscoelastic mechanisms including the sub-Rouse modes and the segmental α-relaxation in polymer thin films, and furthermore the changes can rationalize the enhanced flow at the surface as well as the reduction of glass transition temperature.

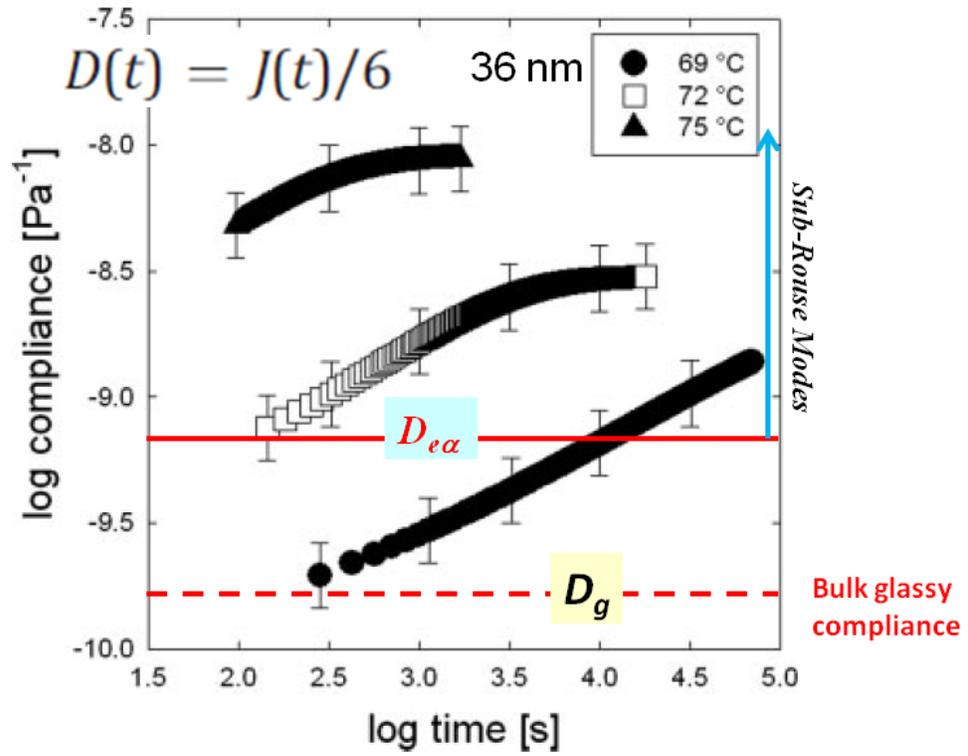

Fig.1. Creep compliance curves for a 36 nm thick PS film at temperatures of 69, 72, and 75.8 C. Data from Ref.[34] redrawn.

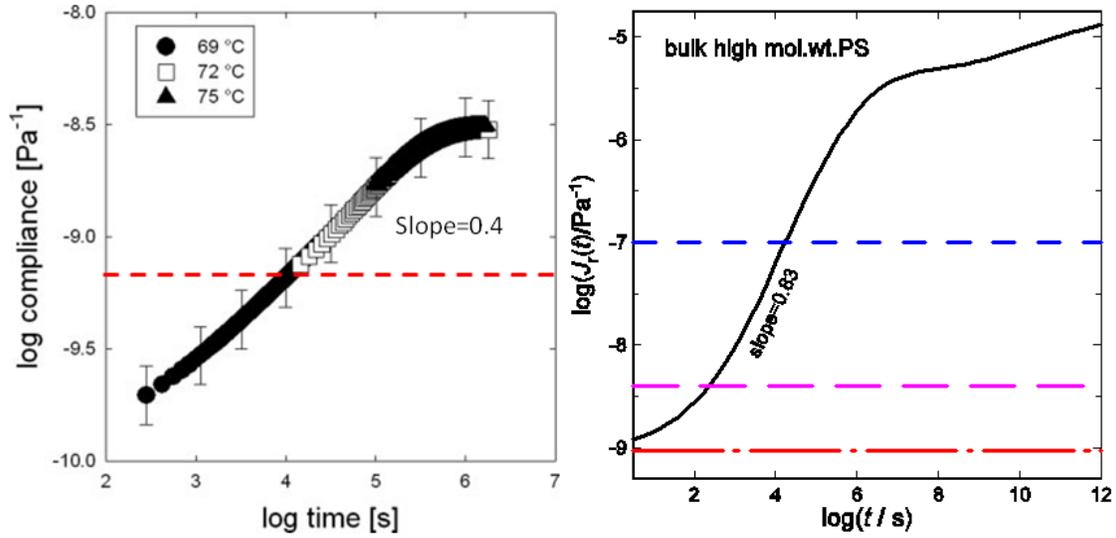

Fig, 2. (a) Master curve constructed from data of the 36 nm thick PS film shown in Fig.1 (data from Ref.[34] are redrawn). (b) Master curve constructed from the recoverable shear creep compliance $J_r(t)$ data of 600,000 Da bulk PS taken at temperatures above $T_g$=100 C by Plazek and O'Rourke Ref.[22]. The horizontal lines show the bounds of the contributions from the segmental α-relaxation and the sub-Rouse modes according to Eqs.(1) and (2).